\documentclass[%
 reprint,
 superscriptaddress,
 amsmath,amssymb,
prl,
]{revtex4-2}

\usepackage{graphicx}
\usepackage{dcolumn}
\usepackage{bm}
\usepackage{chemformula}
\usepackage{hyperref}
\usepackage{amsfonts,amssymb}

\usepackage{tabularx}
\usepackage{amsmath}
\usepackage{float}

\begin{document}


\title{Robust triple-$\mathbf{q}$ magnetic order with trainable spin vorticity in \ch{Na2Co2TeO6}}

\author{Xianghong Jin}
\affiliation{International Center for Quantum Materials, School of Physics, Peking University, Beijing 100871, China}

\author{Mengqiao Geng}
\affiliation{International Center for Quantum Materials, School of Physics, Peking University, Beijing 100871, China}

\author{Fabio Orlandi}
\affiliation{%
ISIS Facility, Rutherford Appleton Laboratory Harwell Oxford, Didcot, OX1 10QX (UK)
}%

\author{Dmitry Khalyavin}%
\affiliation{%
ISIS Facility, Rutherford Appleton Laboratory Harwell Oxford, Didcot, OX1 10QX (UK)
}%

\author{Pascal Manuel}
\affiliation{%
ISIS Facility, Rutherford Appleton Laboratory Harwell Oxford, Didcot, OX1 10QX (UK)
}%

\author{Yang Liu}
\email{liuyang02@pku.edu.cn}
\affiliation{International Center for Quantum Materials, School of Physics, Peking University, Beijing 100871, China}

\author{Yuan Li}
\email{yuan.li@iphy.ac.cn}
\affiliation{International Center for Quantum Materials, School of Physics, Peking University, Beijing 100871, China}
\affiliation{Beijing National Laboratory for Condensed Matter Physics, Institute of Physics, Chinese Academy of Sciences, Beijing 100190, China}

\date{\today}

\begin{abstract}
Recent studies suggest that the candidate Kitaev magnet \ch{Na2Co2TeO6} possesses novel triple-$\mathbf{q}$ magnetic order instead of conventional single-\textbf{q} zigzag order. Here we present dedicated experiments in search for distinct properties expected of the triple-\textbf{q} order, namely, insensitivity of the magnetic domains to weak $C_3$ symmetry-breaking fields and fictitious magnetic fields generated by the spin vorticity. In structurally pristine single crystals, we show that $C_3$ symmetry-breaking in-plane uniaxial strains do not affect the order's magnetic neutron diffraction signals. We further show that \textbf{c}-axis propagating light exhibits large Faraday rotations in the ordered state due to the spin vorticity, the sign of which can be trained via the system's ferrimagnetic moment. These results are in favor of the triple-\textbf{q} order in \ch{Na2Co2TeO6} and reveal its unique emerging behavior.
\end{abstract}

\maketitle

Recently, evidence for a triple-$\mathbf{q}$ magnetic ground state \cite{CWJPRB2021,YWLPRR2023} has been found in the layer cobalt oxide Na$_2$Co$_2$TeO$_6$. This compound has been proposed to approximate the Kitaev honeycomb model \cite{LHMPRB2018,SRPRB2018} with its relatively localized 3$d$ electrons of Co$^{2+}$, which at first sight would allow a simpler model description than the 4$d$ and 5$d$ counterparts. Compared to the single-$\mathbf{q}$ zigzag order, which is widely considered in candidate Kitaev magnets \cite{LXPRB2011,YFPRB2012,SJAPRB2015,JRDPRB2015,HBCPRB2016,LEPRB2016,BAKPRB2017,JQYPRM2019,VOGARXIV2024} due to its proximity to quantum spin liquid phase in simplified models \cite{CJPRL2013,RJGPRL2014}, a triple-$\mathbf{q}$ magnetic ground state appears to be less desired because of its stability requirement of high-order magnetic interactions \cite{CWJPRL2023,WJCPRB2023,PRPRB2023,JPNPJ2024}, which adds complexity to theoretical analyses. But if even a 3$d$ system can be established to have triple-$\mathbf{q}$ magnetic order, the relevance of electron itinerancy may need to be reexamined in a broader range of candidate Kitaev magnets.

Multi-$\mathbf{q}$ magnetic structures are difficult to determine with conventional diffraction experiments alone \cite{KOUVELJPCS1963,KawarazakiPRL1988,MWLJP1990,PBPRB1997,FRSPRB2000}. However, such magnetic structures have two central characteristics that may lead to distinct experimental signatures. Firstly, by preserving the crystallographic rotational symmetries, the multi-$\mathbf{q}$ magnetic order may have substantially fewer orientational domains than the single-$\mathbf{q}$ ones, thus making the domain distribution less sensitive to rotational symmetry-breaking fields. In fact, absence of field training effects has been used as an indication of multi-$\mathbf{q}$ order in a variety of materials \cite{PBPRB1997,KSPRB2000,YWLPRR2023,GYCPRB2024,SHPRB2023}. Secondly, multi-$\mathbf{q}$ magnetic structures are often non-collinear or even non-coplanar. The proposed triple-$\mathbf{q}$ order in \ch{Na2Co2TeO6} features spin vorticity \cite{CWJPRB2021}, which reverses sign under time reversal. As a result, the spin vorticity may act as a fictitious magnetic field even though the order is antiferromagnetic by nature. It may be observed via magnetotransport measurements \cite{SRPRL2001,NAPRL2009,LMPRL2009,TKSCI2019,PPNC2023,THNP2023} and has potential for technological applications such as in spintronics devices.

\begin{figure}[!b]
\centering{\includegraphics[scale=0.7]{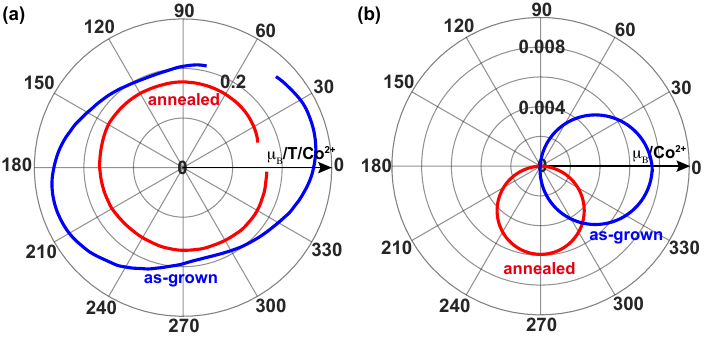}
\caption{\label{fig1}(a) \textbf{ab}-plane angle dependence (\textbf{a} axis at 0$^\circ$) of magnetic susceptibility in the antiferromagnetic state of \ch{Na2Co2TeO6}, measured at 2 K in 0.1 T fields on the same crystal in its as-grown (blue) and annealed (red) states. (b) \textbf{ac}-plane angle dependence (\textbf{a} axis at 0$^\circ$) of ferrimagnetic moment for the same crystal. The crystal was cooled in a tilted field of 0.1 T applied along $+40^\circ$, and measured subsequently at 2 K after switching off the field. }}
\end{figure}

Unfortunately, exploration of the triple-$\mathbf{q}$ magnetic order in \ch{Na2Co2TeO6} along these paths has been limited thus far. On the one hand, while in-plane magnetic fields have been shown to have limited ability to repopulate magnetic domains, the result may be attributed to structural pinning effects that await experimental clarification \cite{YWLPRR2023}. On the other hand, the insulating nature of \ch{Na2Co2TeO6} calls for alternative probes for the spin vorticity not based on electronic transport.

\begin{figure*}[!th]
\begin{centering}
\includegraphics[scale=0.95]{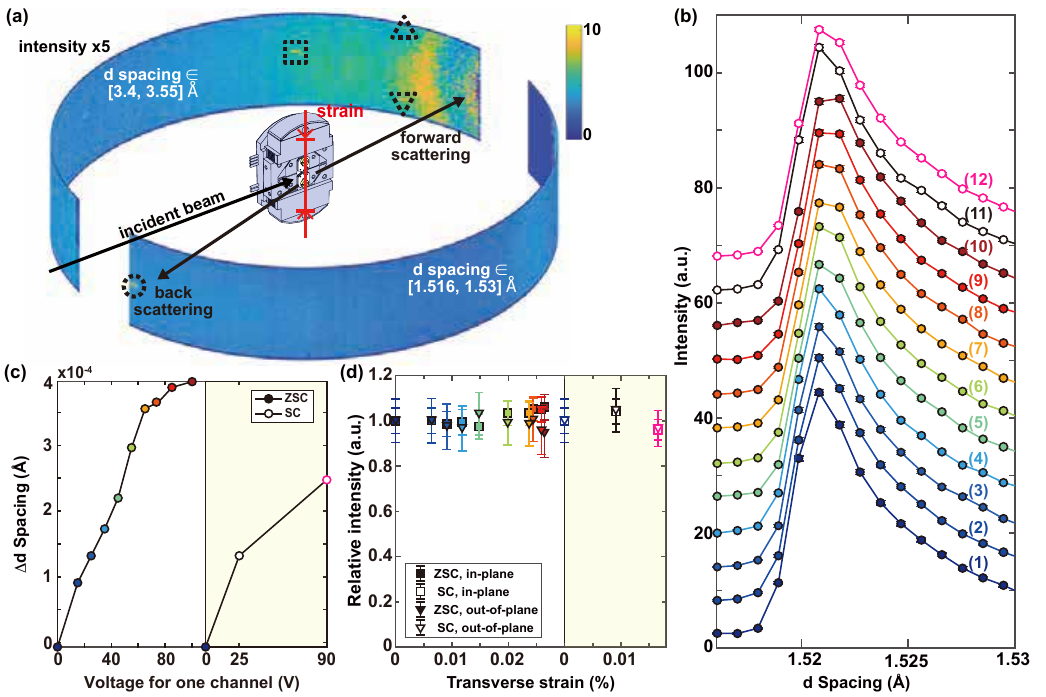}
\caption{\label{fig2}(a) Illustration of scattering geometry. The sample is mounted in a strain cell (enlarged for clarity) for compression in the vertical direction parallel to the crystallographic \textbf{b} axis. Neutrons are scattered primarily in horizontal directions onto the detector banks. Data in two different ranges of $d$-spacing, selected by the neutron time of flight, are displayed on the detector banks. The upper bank measures magnetic diffraction peaks at $\mathbf{q}=(0.5,0,3)$ (dashed square), $(0.5,-0.5,3)$ (dashed triangle), and $(0,0.5,3)$ (dashed inverted triangle), which are shape-coded with data in panel (d). The lower bank measures a structural peak at $(3,0,0)$ (dashed circle) in a back-scattering geometry for optimal momentum resolution. (b-c) Color-coded $d$-spacing profiles of the $(3,0,0)$ peak and the relative shifts as a function of the voltage applied to the piezoelectric stacks of the strain cell. The shifts are used for inference of transverse strains \cite{SM} shown in (d): (1) unstrained, (2-10) strains applied after zero-strain cooling (ZSC), (11-12) strains applied during cooling (SC). (d) Relative intensity variation of two magnetic diffraction peaks versus strain at 11 K. The third peak is weak due to being close to detector edge and thus not shown. }
\end{centering}
\end{figure*}

In this Letter, we present dedicated experiments to test the above expected characteristics of the triple-\textbf{q} order  in \ch{Na2Co2TeO6}. In high-quality single crystals that have minimal residual strain, we use neutron diffraction to demonstrate that the magnetic order is insensitive to in-plane strain, \textit{i.e.}, the strain causes no repopulation of the domains. To work around the system's insulating nature, we use optical Faraday rotation to reveal a large effective magnetization which arises from the spin vorticity in the ordered state. The sign of the effect can be controlled via the system's weak ferrimagnetic moment \cite{YWLPRB2020} by cooling the sample in weak \textbf{c}-axis fields. Our results provide new insight into the symmetry and domain control of the triple-\textbf{q} magnetic order in \ch{Na2Co2TeO6}.

All our measurements were performed on annealed high-quality single crystals of \ch{Na2Co2TeO6}. The annealing process, detailed in \cite{SM}, releases frozen-in strains residual from the growth. To demonstrate its effectiveness, we have measured a crystal at 2 K for the magnetic susceptibility in the \textbf{ab} plane and the ferrimagnetic moment \cite{YWLPRB2020} projection in the \textbf{ac} plane, comparing results obtained before and after annealing [Fig.~\ref{fig1}]. The annealing successfully recovers the structural $C_3$ symmetry in the in-plane responses and aligns the ferrimagnetic net moment along the \textbf{c} axis. These are expected outcome of the triple-\textbf{q} order, which, unlike the zigzag order, preserves the $C_3$ structural symmetry and constrains the ferrimagnetic moment to be parallel to \textbf{c}. The fact that the ferrimagnetic moment can have a non-zero in-plane component (along a seemingly random direction, see Fig.~\ref{figS1}) before the annealing indicates that the sample is not only strained in its as-grown state but also exhibiting strong magneto-elastic coupling. We therefore consider the annealing process a necessary preparatory step for studying the effects of external fields on \ch{Na2Co2TeO6}.

Figure \ref{fig2} presents our neutron diffraction measurements performed with \textit{in situ} in-plane uniaxial strain. The sample is compressed vertically along its \textbf{b} axis. While we cannot directly measure the lattice contraction along \textbf{b} due to the limited neutron detector coverage [Fig.~\ref{fig2}(a)], we do observe a slight lattice expansion in the horizontal direction [Fig.~\ref{fig2}(b-c)]. This reflects the effect of the compression via the Poisson ratio (ranging between 0.16 and 0.25 in typical layered materials \cite{graphitePoisson0.16,bulkMoS20.25,graphene0.16-0.21,hBN0.21}). Based on the data, we estimate that we have applied a maximal compressive strain of about 0.1\% along the \textbf{b} axis. This is a substantial strain for a honeycomb cobalt oxide like \ch{Na2Co2TeO6}. In its sister compound, \ch{Na3Co2SbO6}, a similar inequality in the planar lattice parameters already corresponds to very large magnetic anisotropy \cite{LXTPRX2022}.

\begin{figure}[!b]
\centering{\includegraphics[scale=1.2]{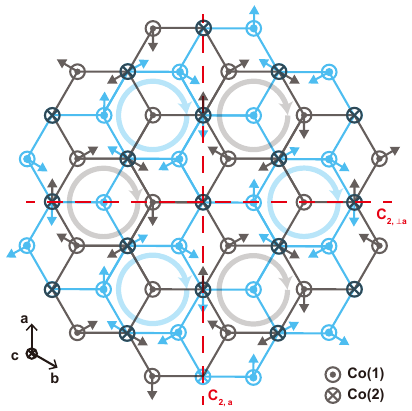}
\caption{\label{fig3}A schematic of triple-\textbf{q} order in two honeycomb Co layers (blue and black). Red dashed lines indicate two types of $C_2$ axes in the $P6_322$ space group, $C_{2, \perp a}$ and $C_{2, a}$, which lie inside and between the Co layers, respectively. Short arrows indicate in-plane spin components. Uniform and opposite signs of \textbf{c}-axis spin components are assumed for each of the symmetry-distinct Co(1) and Co(2) sites. Circular arrows indicate spin vorticity in the layers. The displayed spin configuration maximally preserves the structural rotational symmetries and has uniform spin vorticity (see text).}}
\end{figure}

Under the uniaxial strains, our measurements of the magnetic diffraction peaks [Fig.~\ref{fig2}(d)] at different $M$ points of the Brillouin zone resolve no intensity variation throughout the entire strain range. These $M$ points correspond to different single-\textbf{q} components of the magnetic structure. Similar null results have been obtained after the sample is cooled from the paramagnetic state into the ordered state under strain [Fig.~\ref{fig2}(d)]. We therefore find that, even in a structurally pristine sample where magnetic-domain pinning by the lattice is minimized, a repopulation of magnetic domains does not occur above the $\sim5\%$ level of our statistical uncertainty. This is incompatible with any single-\textbf{q} scenarios of the magnetic order, but fully consistent with the triple-\textbf{q} scenario. In fact, even in the context of the triple-\textbf{q} order, where diffraction intensities at different $M$ points should be understood as components of the same order parameter, their insensitivity to the applied strain is somewhat surprising. It demonstrates that the triple-\textbf{q} order is robustly preserving the $C_3$ symmetry in the magnetic structure, in spite of the lowered lattice symmetry. This in turn suggests that the magnetic in-plane anisotropy of \ch{Na3Co2SbO6} \cite{LXTPRX2022} is caused not merely by the deformed two-dimensional geometry of the honeycomb lattice, but rather by the chemical environment of Co$^{2+}$ in the three-dimensional monoclinic crystal structure, which breaks the local $C_3$ symmetry more strongly. Consistent with this notion, a recent study \cite{DEPRB2023} revealed that the antiferromagnetic order in \ch{Na2Co2TeO6} also sensitively depends on structural polymorphs (hexagonal vs. monoclinic).

Figure~\ref{fig3} presents a schematic of the triple-\textbf{q} magnetic structure of \ch{Na2Co2TeO6} in a Co bilayer, which reflects the fact that all magnetic diffraction peaks are observed at integer momentum indices along $\mathbf{c^*}$ \cite{CWJPRB2021}. Due to the $C_3$ nature of the triple-\textbf{q} order, its order parameter must have equal amplitude at the three \textbf{q}'s. This, combined with ferrimagnetism in the system along \textbf{c} axis \cite{YWLPRB2020}, constrains the magnetic space group to be $P6_32'2'$ in a $\mathrm{2\times2\times1}$ unit cell
\footnote{The magnetic space group can be uniquely constrained as follows. Possible magnetic subgroups of parent space group $P6_322$ compatible with three \textbf{q}s are $P6_32'2'$, $P6_3'22'$, $P6_3'2'2$, $P6_322$, $C2'2'2_1'$, $C22'2_1'$, $C2'2'2_1$, $C222_1$, $P2_1'$ and $P2_1$. Out of these, the followings are compatible with the $C_3$ symmetry preserving order parameter of the triple-\textbf{q} order: $P6_32'2'$, $P6_3'22'$, $P6_3'2'2$ and $P6_322$. Among these magnetic space groups, only $P6_32'2'$ allows for ferrimagnetism along \textbf{c} axis. }. Notably, as the $C_{2,a}$ axes go between adjacent Co layers (Fig.~\ref{fig3}), the combined $C_{2,a}\mathcal{T}$ symmetry interchanges the two layers and enforces the sign of the spin vorticity to be uniform in all layers. One possible definition of the spin vorticity is $\frac{1}{N}\sum_{i} \mathbf{r}_i \times \mathbf{S}_i$, where $\mathbf{S}_i$ and  $\mathbf{r}_i$ are the $i$th spin and its displacement from a vortex center, respectively, and $N$ the number of sites considered. This quantity changes sign under either spatial inversion or time-reversal operations, and it vanishes in the zigzag scenario because time reversal combined with lattice translation remains a symmetry. Due to the lack of spatial inversion symmetry in the crystal structure, the spin vorticity and \textbf{c}-axis ferrimagnetism both act as time-reversal odd axial vectors and transform in the same way under the symmetry operations of the parent space group. Therefore, the two must be coupled to each other. In fact, the only ambiguity in our making of the schematic in Fig.~\ref{fig3} is how the \textbf{c}-axis spin components are drawn with respect to the in-plane components, which is determined by details in the oxygen coordination and spin-orbit coupling of Co \cite{NFarXiv2024}. The bottom line here is that, regardless of how the coupling is realized, we expect it to enable us to control the sign of spin vorticity via the net ferrimagnetic moment, which is trainable by cooling crystals in a weak \textbf{c}-axis magnetic field \cite{YWLPRB2020}.

\begin{figure}[!ht]
\includegraphics[scale=0.78]{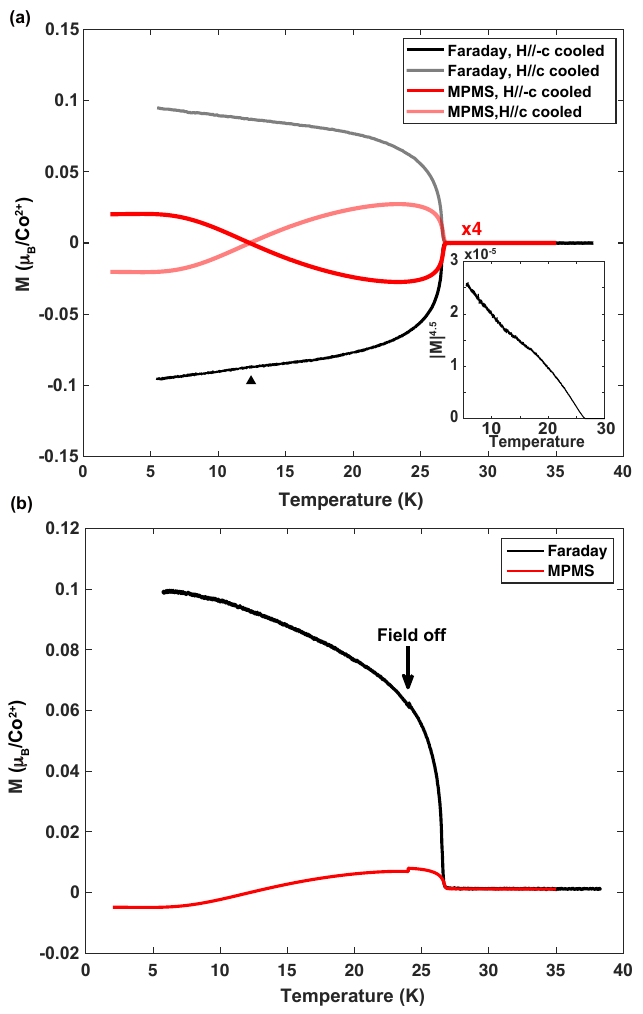}
\caption{\label{fig4}(a) Magnetization inferred from Faraday rotation in comparison to the net ferrimagnetic moment, measured on the same crystal upon warming in zero field, after cooling the sample in 200 Oe magnetic fields applied parallel or anti-parallel to the \textbf{c} axis. The ferrimagnetic moment data are multiplied by a factor of 4 for clarity. Filled triangle indicates a kink in the data (see text), which is more clearly shown in the inset and Fig. \ref{figS4} with magnetization to the power of 4.5 versus temperature. (b) Similar to the data in (a), but measured upon cooling the sample, first in a 200 Oe \textbf{c}-axis field, and then in zero field down to the lowest temperatures. The small decreases in the measured values when field is turned off are due to the disappearance of field-induced magnetization in the sample. }
\end{figure}

Given the insulating and transparent nature of \ch{Na2Co2TeO6}, optical Faraday rotation offers a unique probe of the effective magnetization associated with the spin vorticity. To do this, we have employed an experimental set up based on a reflection-geometry Sagnac interferometer detailed in \cite{SM}. The sample is first measured in its paramagnetic state for both Faraday rotation angle and magnetization (using a Quantum Design MPMS) in a small \textbf{c}-axis magnetic field. This allows us to find the conversion coefficient between Faraday rotation angles and magnetization, which we then use for data conversions in the ordered state. The sample is prepared into a single-domain (concerning the magnetic point group $62'2'$, \textit{e.g.}, the sign of the spin vorticity) ordered state by cooling in weak \textbf{c}-axis magnetic fields.

Figure~\ref{fig4} presents our measurement results in conjunction with data for the net ferrimagnetic moment. As has been reported previously \cite{YWLPRB2020}, the ferrimagnetic moment exhibits a sign reversal at a compensation temperature of about 12.5 K, which reflects the fact that the opposite net \textbf{c}-axis moments of Co(1) and Co(2) grow with decreasing temperature at somewhat different rates. In contrast, the effective magnetization inferred from Faraday rotation has a significantly greater magnitude, and it grows monotonically with decreasing temperature. A tiny kink can be noticed at about 12.5 K in the data in Fig.~\ref{fig4}(a), possibly reflecting the direct contribution of the net magnetization. The kink can be more clearly seen in the inset (and Fig. \ref{figS4} \cite{SM}) as a deviation from the nearly linear increase of $\mathrm{|M|}^{4.5}$ with decreasing temperature below the transition, where the power of 4.5 empirically describes the critical behavior of the order parameter. Importantly, the Faraday rotation at the lowest temperature amounts to an effective magnetization that is over ten times greater than the ferrimagnetic moment, which strongly suggests that it is primarily contributed \textit{not} by the plain magnetic moments, but instead by the spin vorticity in the triple-\textbf{q} ground state. The latter is a robust and uniform property of the entire magnetic structure (Fig.~\ref{fig3}) and does not change sign versus temperature. Indeed, the sign of the vorticity is controlled by that of the ferrimagnetic moment defined just below the ordering temperature: we can switch off the training field at 24 K and still obtain the same training result upon further cooling [Fig.~\ref{fig4}(b)].

Our observation of the spin vorticity with Faraday rotation highlights the power of magneto-optical effects for the determination of complex magnetic order in insulating materials. Indeed, the spin vorticity can be considered a distinct indicator for the triple-\textbf{q} order, as it is not expected to occur in the case of zigzag order. The ``$\mathbf{q}=0$'' (\textit{i.e.}, spatially uniform) character of the spin vorticity can be viewed as a result of the fact that the triple-\textbf{q} order's $\mathbf{q}_1+\mathbf{q}_2+\mathbf{q}_3 =0$, which enables its coupling with, if not mandating by symmetry, the system's ferrimagnetism \cite{YWLPRB2020,NFarXiv2024}. Besides, \ch{Na2Co2TeO6} belongs to a family of compounds in which spin vorticity is controllable with external magnetic field \cite{DLNC2021,PPNPJ2022}. Even in an insulator, spin vorticity might leave distinct signatures in thermal transport experiments sensitive to time-reversal symmetry breaking, \textit{e.g.} anomalous thermal Hall effects.

On a final note,  it is instructive to examine recent report about magnetoelectricity in \ch{Na2Co2TeO6} \cite{ZSZPRB2023}, which has been argued to have the ability to reveal the magnetic order's symmetry \cite{ZSZPRB2023}. Our analysis and data in Figs.~\ref{fig3} and \ref{fig4} pose new requirements on such experiments. In order for a magnetoelectric signal associated with the triple-\textbf{q} order to be experimentally observable (\textit{i.e.}, not washed out by domain average), the sample needs to be prepared not only into a single-domain triple-\textbf{q} state, but also with the two classes of combined $C_2\mathcal{T}$ symmetries broken. This requires cooling the sample in a weak \textbf{c}-axis field, followed by measurements of electric polarization in the ordered state with a sizable in-plane magnetic field. To our knowledge, such dedicated experiments have not been reported so far.

In conclusion, our new measurement results unequivocally support the presence of triple-\textbf{q} order in \ch{Na2Co2TeO6}. Using structurally pristine single crystals subject to symmetry-breaking training fields, we establish both the order's preserved $C_3$ symmetry about the \textbf{c} axis and its emergent spin vorticity uniquely coupled to $\mathbf{q}=0$ ferrimagnetism, thus providing the last two missing pieces of evidence expected of the triple-\textbf{q} order. The confirmed multi-\textbf{q} nature of the magnetic ground state in a $3d$-electron candidate Kitaev material, as well as our experimental methodology in uncovering the distinct characteristics of the order, brings valuable insight and considerations for deepening our understanding of frustrated magnetism in related insulating quantum materials.

\begin{acknowledgments}
We are grateful for technical support by Jianping Sun and for discussions with Riccardo Comin, Mingquan He, Lukas Janssen, and Tao Lu. The work at Peking University was supported by the National Basic Research Program of China (Grant No. 2021YFA1401900) and the National Natural Science Foundation of China (Grant Nos. 12061131004 and 11888101). The neutron scattering experiment was performed at the ISIS facility in the United Kingdom, under a user program (No. 2310181) and the raw data are available at \cite{WISH2023}.
\end{acknowledgments}


\bibliography{ref}

\pagebreak
\pagebreak

\widetext
\begin{center}
\textbf{\large Supplemental Material for \\``Robust triple-$\mathbf{q}$ magnetic order with trainable spin vorticity in \ch{Na2Co2TeO6}''}
\end{center}
\setcounter{equation}{0}
\setcounter{figure}{0}
\setcounter{table}{0}
\makeatletter
\renewcommand{\theequation}{S\arabic{equation}}
\renewcommand{\thetable}{S\arabic{table}}
\renewcommand{\thefigure}{S\arabic{figure}}
\renewcommand{\bibnumfmt}[1]{[S#1]}

\section{Single crystal preparation}
 Single crystals of \ch{Na2Co2TeO6} were synthesized with a flux method \cite{YWLPRL2022}. For annealing after growth, single crystals were put into the crucible and then a box furnace. The furnace was heated up to 600 $^\circ$C in 5 hours and maintained at 600 $^\circ$C for 1 hours. Then it was cooled with 6 $^\circ$C per hour to 20 $^\circ$C. The annealing process reduces stacking fault and restores $C_3$ symmetry. The sample for strain experiment was cut to a needle of dimension 3.5 \ch{mm} $\times$ 0.8 \ch{mm} $\times$ 0.3 \ch{mm} (\textbf{b} $\times$ \textbf{\ch{b^*}} $\times$ \textbf{c}).

\section{Magnetic susceptibility measurement}
DC magnetometry in Fig. \ref{fig1} and Fig. \ref{fig4} was performed with a Quantum Design MPMS equipped with a sample rotator and brass sample holder respectively. The magnetization of \ch{Na2Co2TeO6} consists of a regular, susceptibility-related component and a ferrimagnetic component, i.e. $M=M_\mathrm{AFM}+M_\mathrm{ferri}=\chi_\mathrm{AFM}*H+M_\mathrm{ferri}$ \cite{YWLPRB2020}. The angle dependence of both \textbf{ab}-plane and \textbf{ac}-plane ferrimagnetism $M_\mathrm{ferri}$ was measured without field on after cooling to 2 K with 0.1 T. Then the angle dependence of \textbf{ab}-plane magnetization $M$ was measured with 0.1 T field on after cooling to 2 K with 0.1 T along same direction as training \textbf{ab}-plane ferrimagnetism. The angle dependence of AFM was obtained with $M-M_\mathrm{ferri}$. 0.1 T is chosen due to the following reasons: (1) 0.1 T along nearly any direction except \textbf{a}, \textbf{b} or \textbf{c} is enough for training a saturate ferrimagnetism \cite{YWLPRB2020} and (2) AFM part contributes linear response to magnetic field under 0.1 T \cite{YWLPRB2020}. Therefore, isotropic response in Fig. \ref{fig1}(a) at 0.1 T is compatible with $C_3$ symmetry rather than $C_2$.

The ferrimagnetism in Fig. \ref{fig4}(a) is measured during warming up with zero field, before which it is FC trained till base temperature with 200 Oe field along \textbf{c} or \textbf{-c} direction, while in Figure \ref{fig4}(b) it is FC trained till 24 K with 200 Oe field along \textbf{c} axis and then cooled down without field.

\section{Neutron diffraction experiment and strain estimation}
The experiment was performed on a cold neutron time-of-flight diffractometer, WISH (Wide angle In a Single Histogram), at the ISIS facility in the United Kingdom, using a strain cell CS200t as the sample environment \cite{WISH,HCWRSI2014}. An annealed sample on the strain cell was mounted with reciprocal vector (H, 0, L) in the horizontal plane and the uniaxial strain was along \textbf{b} axis. The results in Fig. \ref{fig2} was measured at 11 K. The SC data was measured after cooling with strain from 35 K and ZSC data was measured following an initial zero-strain cooling from 300 K.

The raw data was processed with the software, Mantid (Manipulation and Analysis Toolkit for Instrument Data) \cite{Mantid1,Mantid2}, which corrects the white beam intensity profile versus energy but not for the Lorentz factor. The Lorentz factor is proportional to $d^2\lambda^2$, where d is the spacing of adjacent layers in real space and $\lambda$ is the wavelength of incident neutron. Since the incident wavelength are approximately 3.2 \ch{\AA} and 2.0 \ch{\AA} respectively, the intensities of in-plane and out-of-plane magnetic Bragg peaks are consistent with Lorentz factor within experimental accuracy.

The strain is controlled via voltage on the piezoelectric stacks. As the correspondence between strain and voltage depends on temperature, shifts of center of Bragg peak at 11 K [Fig. \ref{fig2}(b-c)]  were used to estimate the real strain on the sample. Due to the scattering geometry, Bragg peaks along the strain direction is unable to detect. Therefore, we estimated the transverse strain through the shift of the center of an in-plane structural Bragg peak \textbf{q}=(3, 0, 0) at 11 K, which was obtained with
$$\mathrm{center}=\frac{\sum_i d_i*y_i}{\sum_i d_i}$$ where $d_i$ stands for d spacing of the data within [1.51, 1.56] \ch{\AA} and $y_i$ is corresponding intensity. The zero center shift is set as zero applied voltage, which was identical to another annealed free-standing sample measured on WISH. The estimated transverse strain together with measurement conditions is detailed in \ref{table1}, with numbers corresponding to those in Fig. \ref{fig2}(b). Combined with Poisson ratio of usual $2D$ material 0.16-0.25 \cite{graphitePoisson0.16,bulkMoS20.25,graphene0.16-0.21,hBN0.21}, strain along \textbf{b} axis can be estimated.

\section{Optical Faraday measurement and conversion to effective magnetic moment}

The optical Faraday measurements were carried out using an all-fibre zero-loop Sagnac interferometer (SI) similar to \cite{JXPRL2006,KAPITULNIK,HDQCPL2024}, installed in a closed-cycle cryostat with 4.5 K base temperature. We use an ASE (amplification of spontaneous emission) broadband light source centered at wavelength 1550 nm whose width is 40 nm, which is essential for only TR symmetry-breaking Faraday effect measurement \cite{KAPITULNIK}. The radius of light spot on the sample is about 100 \ch{$\mu$}m. The probing light is along \textbf{c} axis. The sample is glued to silicon wafer deposited with 100 nm gold on top with very thin GE Varnish, enabling Faraday detection with SI.

In Fig. \ref{fig4}(a), the sample is trained with 200 Oe field along \textbf{c} or \textbf{-c} direction from above $T_\mathrm{N}$ to base temperature. After training, optical Faraday rotation was measured during warming up with zero field. And in Fig. \ref{fig4}(b), it was measured with 200 Oe field along \textbf{c} direction from above $T_\mathrm{N}$ to 24 K and then to base temperature without field.

The sign of Faraday rotation is defined with relative sign to that in paramagnetic state, which is set to be the same as direction of external magnetic field.

The Faraday rotation angle is converted to effective magnetic moment with the following equation, $$\Theta_F=VBd=\mu_0Vd(H+M)$$ where $\Theta_F$ is Faraday rotation, $d$ the thickness of sample, $H$ the magnetic field, $M$ the magnetization and $V$ the Verdet constant. The Verdet constant is obtained with the knowledge of Faraday rotation, sample thickness and magnetic susceptibility at 35K. With the Verdet constant, Faraday rotation is converted to effective magnetic moment. The giant effective magnetic moment in the antiferromagnetic state compared with ferrimagnetism in MPMS measurement suggests existence of fictitious field due to spin vorticity.

\section{Supplementary Figures and Table}

\begin{figure}[h]
\centering{\includegraphics[scale=0.7]{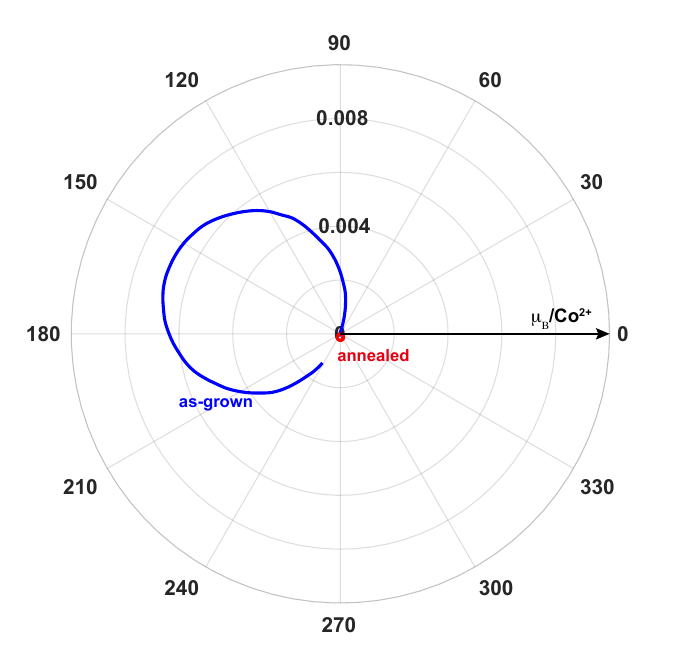}}
\caption{\textbf{ab}-plane angle dependence (\textbf{a} axis at 0$^\circ$) of ferrimagnetic moment for the same crystal as in Fig. \ref{fig1}. The sample was cooled in a tilted field of 0.1 T applied along +145$^\circ$ and measured subsequently at 2 K after switching off the field.}
\label{figS1}
\end{figure}

\begin{figure}[h]
\centering{\includegraphics[scale=0.8]{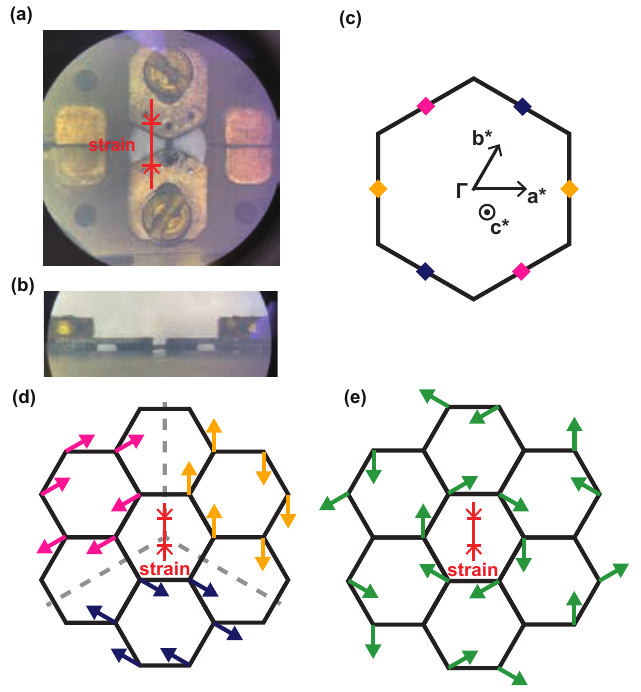}}
\caption{(a-b) Top and side views of the strain cell loaded with annealed \ch{Na_2Co_2TeO_6} single crystal with strain along \textbf{b} axis. (c) $2D$ reciprocal space and illustration of magnetic Bragg peaks, where yellow, cyan, dark blue dots correspond to magnetic Bragg peaks indicated by dashed square, dashed triangle, dashed inverted triangle in Fig. \ref{fig2}(a). (d-e) Demonstration of zigzag order with $C_3$-related domains and triple-$\mathbf{q}$ order in \textbf{ab} plane. The yellow, cyan, dark blue domains divided by grey dash lines correspond to magnetic Bragg peaks in panel (c) for 3 domains respectively. And correspondence of color and shape[Fig. \ref{fig2}(a)] is the same as that in panel (c). And magnetic Bragg peaks of triple-\textbf{q} order consist of all peaks in (c). }
\label{figS2}
\end{figure}

\begin{table}[ht]
\caption{\label{table1}
Measurement conditions used in Fig.~\ref{fig2}(b).
}
\begin{ruledtabular}
\begin{tabular}{ccc}
\textrm{Number}&
\textrm{Transverse strain}&
\textrm{ZSC/SC}\\
\colrule
(1) & 0\% & ZSC \\
(2) & 0.0064\% & ZSC \\
(3) & 0.0091\% & ZSC \\
(4) & 0.0118\% & ZSC \\
(5) & 0.0148\% & ZSC \\
(6) & 0.0199\% & ZSC \\
(7) & 0.0237\% & ZSC \\
(8) & 0.0244\% & ZSC \\
(9) & 0.0259\% & ZSC \\
(10) & 0.0264\% & ZSC \\
(11) & 0.0091\% & SC \\
(12) & 0.0166\% & SC \\
\end{tabular}
\end{ruledtabular}
\end{table}

\begin{figure}[h]
\centering{\includegraphics[scale=0.8]{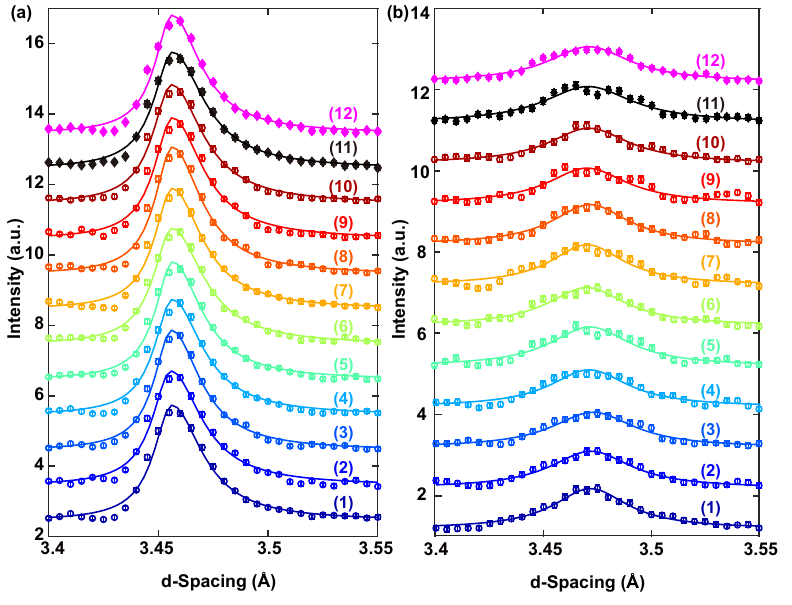}}
\caption{(a-b) Raw data of intensity of M point in plane and out of plane raw data at 11 K respectively, fitted with asymmetric Gaussian profile. The labeled numbers are for inference of transverse strain in \ref{table1}}
\label{figS3}
\end{figure}

\begin{figure}[h]
\centering{\includegraphics[scale=1.0]{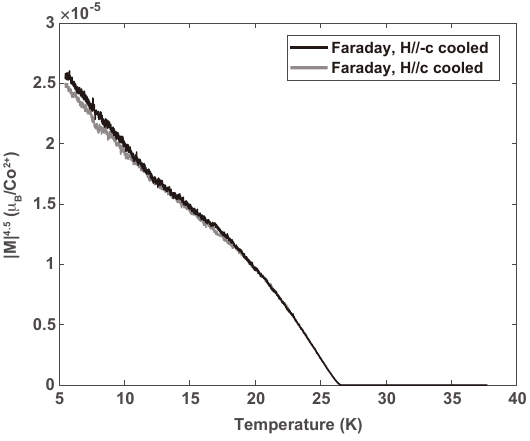}}
\caption{Magnitude of magnetization to the power of 4.5 versus temperature. Same as the data in the inset of Fig. \ref{fig4}(a), a kink at about 12.5 K is clearly seen in both traces as deviation from linear temperature dependence of $\mathrm{|M|}^{4.5}$, which describes the critical behavior of the order parameter (proportional to the spin vorticity).}
\label{figS4}
\end{figure}

\end{document}